%% file: main.tex
\documentclass{edm_template}

\usepackage{color}
\usepackage[normalem]{ulem}
\usepackage{pgfplots}
\usepgfplotslibrary{groupplots}
\usepackage{enumitem}
\usepackage{booktabs}

\pgfplotsset{compat=1.14}

\pgfplotsset{
    colormap={pgfplots}{rgb255=(0,0,255) rgb255=(0,255,255) rgb255=(0,255,29) rgb255=(255,255,0) rgb255=(255,144,0) rgb255=(255,0,0)},
    colorbar style={
            yticklabel style={
                /pgf/number format/.cd,
                fixed,
                precision=1,
                fixed zerofill,
            },
        },
}

\newcommand{\cluster}[1]{C_{#1}}
\newcommand{\feature}[1]{f_{#1}}

\newcommand{\mat}[1]{\mathbf{#1}}
\newcommand{\transpose}[1]{#1^T}
\newcommand{\inverse}[1]{#1^{-1}}
\newcommand{\fnorm}[1]{||#1||_F}
\newcommand{\matmult}{\odot}

\newcommand{\N}{\mathbb{N}}
\newcommand{\R}{\mathbb{R}}

\begin{document}

\title{Tracking Behavioral Patterns among Students in an Online Educational System}

\numberofauthors{3} 
\author{
\alignauthor
Stephan Lorenzen\\
       \affaddr{University of Copenhagen}\\
       \email{lorenzen@di.ku.dk}
% 2nd. author
\alignauthor
Niklas Hjuler \\
       \affaddr{University of Copenhagen}\\
       \email{hjuler@di.ku.dk}
\alignauthor
Stephen Alstrup \\
       \affaddr{University of Copenhagen}\\
       \email{s.alstrup@di.ku.dk}
}

%\additionalauthors{Additional authors: John Smith (The Th{\o}rv{\"a}ld Group,
%email: {\texttt{jsmith@affiliation.org}}) and Julius P.~Kumquat
%(The Kumquat Consortium, email: {\texttt{jpkumquat@consortium.net}}).}
%\date{December 2017}

\maketitle
\begin{abstract}
Analysis of log data generated by online educational systems is an essential task to better the educational systems and increase our understanding of how students learn. In this study we investigate previously unseen data from Clio Online, the largest provider 
of digital learning content for primary schools in Denmark.
We consider data for 14,810 students with 3 million sessions in the period 2015-2017. We analyze student activity in periods of one week. 
By using non-negative matrix factorization techniques, we obtain soft clusterings, revealing
dependencies among time of day, subject, activity type,
activity complexity (measured by Bloom's taxonomy), and performance.
Furthermore, our method allows for tracking behavioral changes of individual students over time, as well as general behavioral changes in the educational system.
Based on the results, we give suggestions for behavioral changes, in order to optimize the learning experience and improve performance.  
\end{abstract}


\section{Introduction + Related work}
How students behave in educational systems is an important topic in educational data mining. Knowledge of this behavior in an educational system can help us understand how students learn, and help guide the development for optimal learning based on actual use. This behaviour can be understood both through an explicit study \cite{DBLP:conf/edm/HuttMWDD16}, or as in this paper through the automatically generated log data of the system.

The analysis of log data is usually done as an unsupervised clustering of students \cite{conf/edm/FauconKD16,DBLP:conf/edm/GelmanRDVJ16,DBLP:journals/corr/abs-1708-04164,kli16b}. A popular approach is to extract action sequences and transform them into an aggregated representation using Markov models \cite{DBLP:journals/corr/abs-1708-04164,kli16b}. The Markov chains can then be clustered by different methods. Klingler et al. did student modeling with the use of explicit Markov chains and the clustering with different distance measures defined on the Markov chains \cite{kli16b}. Hansen et al. assumed the actions sequences to be generated by a mixture of Markov chains and used an heuristic algorithm to find the generating Markov chains \cite{DBLP:journals/corr/abs-1708-04164}. Gelman et al. used non-negative matrix factorization to find clusters for different measures of activity aggregated in weekly periods during a MOOC course. These clusters are then matched from week to week by cosine similarity.

Our work is similar to Gelman et al. \cite{DBLP:conf/edm/GelmanRDVJ16} in that we also use \emph{Non-negative Matrix Factorization} (NMF) to make a soft clustering at the student level in a given time period, however our clustering is only made once, and we are looking at primary school data over a vastly longer period of time, (2 years compared to 14 weeks). 

Our soft clustering by non-negative matrix factorization is based on log data from Clio Online.\footnote{This data is proprietary and not publicly available.}
Clio Online is the largest provider of digital learning for all subjects in the Danish primary school (except mathematics), having 90\% of all primary schools in Denmark as customers.

Using NMF, we assume that the set of features chosen can be represented by a set of fewer underlying behaviors. These underlying behaviours would each be represented  by a cluster in the non-negative matrix factorization. Each student will then get a number for each cluster in each time period representing how much of that underlying behavior he has shown in the given time period. Non-negativity gives the behaviors an additive structure, which is more natural than showing a negative amount of a given behavior.
We reason that the soft clustering will show both the behaviors of individual students, as well as how the behaviors change over time, both individually and on a system-wide level.

In this paper, we will consider two main questions: a) how does student activity in the system affect performance, and b) how does student activity distribute between different levels of Bloom's taxonomy in different subjects. Both questions are important in regards to optimizing learning; the first in relation to performance, the latter in relation to utilization of all taxonomy levels.
\newpage
\section{Experimental Setup}\label{sec:setup}
This section describes our experimental setup and methods. We start by describing our data and how it is preprocessed, and then move on to describing our clustering method.

\subsection{Data Preprocessing}\label{sec:data}
As mentioned, we consider log data generated in the Danish online educational system Clio Online. The system is used in Danish primary schools and contains learning objects across all Danish subjects (except mathematics), for instance texts, videos, sound clips and exercises. Furthermore, the system includes a large number of quizzes, used for evaluating students. Students may use the system for self study, but they may also be assigned homework by their teacher. Our data covers 14,810 students.

The raw data consists of logs detailing page accesses for individual students in the system. For quizzes, the final score (between 0 and 1) and total time spent for the quiz is also available. In our preprocessing, we combine these log entries to \emph{sessions}. Two consecutive entries are considered in the same session, if they have the same subject, and their timestamps differ by less than some threshold. For our study, we choose this threshold to be 600 seconds, based on recommendations from Clio Online, who have a deeper knowledge of the content and flow of the system (e.g. expected time per page). Furthermore, quizzes are considered separate sessions. A total of 3 million sessions is obtained in this way.

With the sessions defined, we consider student activity in \emph{activity periods}, with a length of one week. The data spans a total of 112 activity periods, starting January 2015 and ending in March 2017. For each activity period, we add an entry for a student, if the student is active (accesses the system) within that period. The entry for the given student contains all sessions for that student, which starts within the activity period. We end up with approximately 677,000 student entries across the 112 periods.
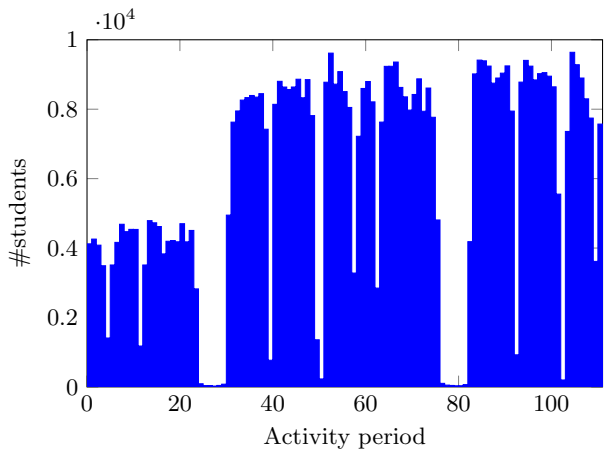
\begin{figure}
    \centering
    \input{figs/activity.tex}
    \caption{Number of students active in each period. Note that period 0 starts on 2015-01-08, while period 111 ends on 2017-03-01. The drops in activity occur due to vacation in Danish primary school, with the two large drops around periods 25 and 79 being due to the summer vacation.}
    \label{fig:student-activity}
\end{figure}
Figure~\ref{fig:student-activity} shows the active number of students in each period. Note the drop in active students around periods 25 and 79; these drops in activity occur due to summer vacation.

The final step of data preprocessing is the feature extraction. For each activity period, a set of activity/performance related features are extracted. The features are chosen so as to answer the questions posed in the previous section.
\begin{table*}
    \centering
    \begin{tabular}{c|l|c|c|c}
        \toprule
        $i$ & $\feature{i}$ & Max & Mean & Variance \\
        \midrule
        1  & Hours between 8AM and 4PM            & 31.85  & 0.940 & 0.862 \\
        2  & Hours before 8AM and after 4PM       & 71.84  & 0.174 & 0.283 \\
        3  & Hours doing exercises                & 3.61   & 0.048 & 0.019 \\
        4  & Hours reading texts                  & 7.73   & 0.344 & 0.148 \\
        5  & Hours taking quizzes                 & 23.76  & 0.231 & 0.297 \\
        6  & Hours working with language subjects & 58.28  & 0.531 & 0.693 \\
        7  & Hours working with societal subjects & 45.96  & 0.294 & 0.285 \\
        8  & Hours working with science subjects  & 103.69 & 0.277 & 0.326 \\
        9  & Average session length in hours      & 7.91   & 0.268 & 0.027 \\
        10 & Average quiz score (in $[0,1]$)      & 1.00   & 0.733 & 0.034 \\
        11 & Hours working with Bloom level 1     & 2.83   & 0.016 & 0.006 \\
        12 & Hours working with Bloom level 2     & 1.64   & 0.008 & 0.002 \\
        13 & Hours working with Bloom level 3     & 1.51   & 0.014 & 0.003 \\
        14 & Hours working with Bloom level 4     & 2.04   & 0.009 & 0.003 \\
        \bottomrule
    \end{tabular}
    \caption{Overview of features.}
    \label{tab:feature-overview}
\end{table*}
A complete overview of all features considered in our experiments is given in Table~\ref{tab:feature-overview}, including the maximum, mean and variance across all active students in all periods. Not all features are used for each experiment, see section~\ref{sec:results}.

All features are aggregates over the activity period. Below follows a detailed description:
\begin{itemize}[noitemsep]
%\item $\feature{1}$ is a measure of the total activity in the system.
\item $\feature{1}$ describes the activity during the period of day, where Danish students are normally in school, while $\feature{2}$ describes the activity during non-school hours.
\item $\feature{3}$, $\feature{4}$ and $\feature{5}$ describe time spent doing exercises, reading texts and taking quizzes respectively.
\item $\feature{6}$, $\feature{7}$ and $\feature{8}$ describe time spent working with different topics: languages (Danish, English, German), societal (social studies, history, etc.) and science (physics, biology, etc.), respectively.
\item $\feature{9}$ is the average session length during the activity period.
\item $\feature{10}$ is the average quiz score; this feature may be missing, if a student takes no quizzes during an activity period, but our analysis methods can handle this, see section~\ref{sec:nmf}.
\item $\feature{11}$, $\feature{12}$, $\feature{13}$ and $\feature{14}$ describe the time spent doing exercises of different complexity, measured by their level in Bloom's taxonomy. We regroup the levels of Bloom's taxonomy into 4 levels:
\begin{enumerate}[noitemsep]
    \item[$\feature{11}$] \textbf{Remember/Understand}: Exercises involving reading and describing, e.g. "Read a map".
    \item[$\feature{12}$] \textbf{Apply}: Exercises involving application of previously learned concepts, e.g. "Practice adjectives".
    \item[$\feature{13}$] \textbf{Analyze/Evaluate}: Exercises involving discussion, analysis and experimenting, e.g. "Work with the poem", "Analyze the game".
    \item[$\feature{14}$] \textbf{Create}: Exercises involving creation of a product, e.g. "Create a cartoon", "Write a story".
\end{enumerate}
\end{itemize} 
Having extracted $m$ features for each student in each period, we construct the matrix $\mat{X}\in\R^{n\times m}$, where each of the $n$ rows consists of the feature vector for an active student in a given activity period. Thus each student occurs several times in $\mat{X}$; once for each period, where they are active.

\subsection{Soft Clustering using Non-negative Matrix Factorization}\label{sec:nmf}
We will utilize non-negative matrix factorization for our soft clustering. The use of NMF as a soft clustering technique has become popular in recent times \cite{Li13}, with applications within several fields, such as clustering of images and documents \cite{nmf-image-clustering,nmf-doc-clustering}. NMF has also seen success in the educational data mining community, for clustering tasks, as well as other tasks such as performance prediction \cite{DBLP:conf/edm/GelmanRDVJ16,ECEL-nmf}. 

NMF is a dimensionality reduction method, in which we are given a non-negative matrix $\mat{X}\in\R^{n\times m}_+$ and $k\in \N$, and wish to determine $\mat{U}\in\R_+^{n\times k},\mat{V}\in\R_+^{k\times m}$, such that $\mat{X}\simeq \mat{U}\mat{V}$. More specifically, we search for $\mat{U}$ and $\mat{V}$, such that the error $\fnorm{\mat{X}-\mat{U}\mat{V}}$ is minimized, where $\fnorm{\cdot}$ is the Frobenious norm.
For our analysis, we need to be able to handle missing values in $\mat{X}$. In this case the NMF problem is reformulated as the \emph{weighted non-negative matrix factorization}, in which we are also given a binary weight matrix $\mat{W} \in \left\{0,1\right\}^{n\times m}$, where a 0 indicates missing data. Now, we wish to find $\mat{U},\mat{V}$ such that $\fnorm{\mat{W}\matmult \left(\mat{X}-\mat{U}\mat{V}\right)}$ is minimized\footnote{$\matmult$ denotes the Hadamard product (element-wise multiplication).}.

\begin{figure}
    \centering
    \input{figs/nmf-cluster.tex}
    \caption{The soft clustering given by NMF.}
    \label{fig:soft-cluster-nmf}
\end{figure}
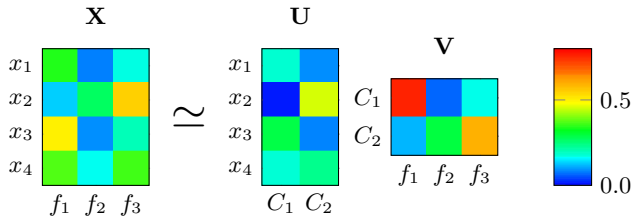
$\mat{U}$ and $\mat{V}$ admits a soft $k$-clustering as shown in Figure~\ref{fig:soft-cluster-nmf}; $\mat{V}$ describes the importance of each feature for each cluster (for instance, $\feature{1}$ has high importance in $\cluster{1}$), while $\mat{U}$ describes the membership of each data point to the different clusters (for instance, $x_3$ is mostly in $\cluster{1}$, while $x_4$ is in both clusters).

Note, that for NMF, we have $\mat{X} \simeq \mat{U}\mat{V} = \mat{U}\mat{I}\mat{V} = \mat{U}\inverse{\mat{A}}\mat{A}\mat{V}$, where $\mat{I}$ is the $k\times k$ identity matrix and $\mat{A}$ is a $k\times k$ invertible matrix. This means that we may rescale $\mat{U}$ and $\mat{V}$ by this matrix, $\mat{A}$, and its inverse. In our analysis, we use this to rescale $\mat{V}$, such that all rows of $\mat{V}$ (the clusters) sum to one, thus making the clusters comparable, and membership of the different clusters easier interpretable.

There exist several algorithms for obtaining the non-negative matrix factorization of $\mat{X}$, for instance basic gradient descent, multiplicative update rules and alternating least squares; \cite{nmf-overview} gives a good overview in the non-weighted setting. Several of these algorithms have been adapted for the WNMF case, while approaches based on \emph{expectation maximization} have also been proposed, see \cite{wnmf-overview}. For our analysis, we will use the weighted version of the multiplicative update method, proposed by Lee and Seung \cite{Lee00}.

The NMF algorithm given in \cite{Lee00}, adopted to WNMF \cite{wnmf-overview}, is as follows:
\begin{enumerate}[noitemsep]
    \item Initialize $\mat{U}$ and $\mat{V}$.
    \item Repeatedly update $\mat{U}$ and $\mat{V}$ by the following rules:
    \begin{align*}
        \mat{U} &\leftarrow \mat{U} \matmult \frac{ \left( \mat{W}\matmult \mat{X} \right) \transpose{\mat{V}} }{
                                                    \left( \mat{W}\matmult \left( \mat{U}\mat{V} \right) \right) \transpose{\mat{V}} } \\
        \mat{V} &\leftarrow \mat{V} \matmult \frac{ \transpose{ \mat{U} } \left( \mat{W}\matmult \mat{X} \right) }{
                                                    \transpose{ \mat{U} } \left( \mat{W}\matmult \left( \mat{U}\mat{V} \right) \right) }
    \end{align*}
    where division is done element-wise.
\end{enumerate}
The literature explores several ways of initializing $\mat{U}$ and $\mat{V}$; in our case, we will simply use random initialization. The alternating optimization steps are applied until the decrease in error reaches below a set threshold.
Finally, Lin has noted that the procedure described above may not converge to a stationary point, hence we modify the update rules as proposed by them \cite{Lin07}.
Furthermore, since we in our case know all missing values of $\mat{X}$ to be bounded by a constant $c$, we modify the above procedure such that 0-weight values of $\mat{U}\mat{V}$ that deviate above $c$ are penalized, i.e. whenever a value $(\mat{U}\mat{V})_{ij}$ with $\mat{W}_{ij} = 0$ gets larger than $c$, we set $\mat{X}_{ij} = c$ and $\mat{W}_{ij} = 1$, before the next update step. If $(\mat{U}\mat{V})_{ij}$ decreases below $c$ again, the weight is reset to 0. 

It remains to be seen, how we select the number of clusters, $k$. For each experiment, we construct clusterings with $k=1,2,...$, and stop when the decrease in error going from $k$ clusters to $k+1$ clusters is below some threshold, which depends on the initial error. As a consequence clusters will be uncorrelated on a student level, since otherwise we would pick a lower $k$.

\section{Experiments and Results}\label{sec:results}
In this section, we present two different experiments using the setup described above. In the first experiment, we investigate the relation between activity, activity type, subject, time of day, average session length and performance. In the second experiment, we investigate the relation between complexities of exercises and subjects. 

\subsection{Performance and Optimal Behavior}
\begin{figure}[t]
    \centering
    \input{figs/e1-V.tex}
    \caption{The cluster matrix for the first experiment.}
    \label{fig:experiment1-V}
\end{figure}
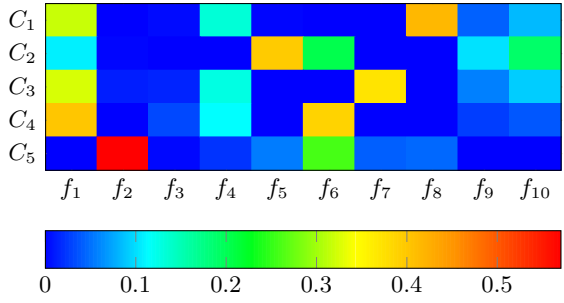

In the first experiment, we investigate the relation between activity, activity type, subject, time of day, average session length and performance, i.e. we consider features $\feature{1},...,\feature{10}$. The features are extracted and $k=5$ is selected, as described in section~\ref{sec:setup}. We run the WNMF algorithm, and obtain the cluster matrix $V$ as shown in Figure~\ref{fig:experiment1-V}.
From the figure, we can make several observations about the clusters:
\begin{itemize}[noitemsep]
%% Cluster 1
\item[$\cluster{1}$] In this cluster, we find students mostly working with the science subjects ($\feature{8}$). These students seem to work mostly during school hours ($\feature{1}$).
%, but there is also some activity at home ($\feature{2}$).
The students also seem to spent a lot of time reading ($\feature{4}$).
%% Cluster 2
\item[$\cluster{2}$] Students in this cluster spend a lot of time taking quizzes ($\feature{5}$). They will spend some time during school hours ($\feature{1}$) and some time working with language subjects ($\feature{6}$). Furthermore, students in this cluster seem to both have fairly long average session length and high performance ($\feature{9}$ and $\feature{10}$).
%% Cluster 3
\item[$\cluster{3}$] In cluster $\cluster{3}$, we see students working with societal subjects ($\feature{7}$). They work during school hours ($\feature{1}$) and spend time reading texts in the system ($\feature{4}$).
%% Cluster 4
\item[$\cluster{4}$] This cluster shows a relationship between being active in school ($\feature{1}$) and spending time in the language subjects ($\feature{6}$). Students in this cluster also spend time reading texts ($\feature{4}$) and doing some exercises ($\feature{3}$).
%% Cluster 5
\item[$\cluster{5}$] The most important feature for $\cluster{5}$ is $\feature{2}$, i.e. the students in this cluster spend most time using the system during non-school hours. These students spent time in all subjects, but mostly languages ($\feature{6}$), and they spent time taking quizzes ($\feature{5}$).
\end{itemize}
From the clusters, we can see that the impact on performance from different behaviors depends on the subject. From cluster $\cluster{2}$, we see that students working mostly with language subjects gain most performance from spending time taking quizzes and working during school hours, whereas students working mostly with societal (cluster $\cluster{3}$) and science (cluster $\cluster{1}$) subjects gain most from reading texts, while working
mostly during school hours. Note that cluster $\cluster{4}$ indicates that students working with languages may also improve performance by reading texts, but to a lesser degree than students working in other subjects. Finally, $\cluster{5}$ indicates that working mostly from home and primarily taking quizzes, does not improve performance. While $\cluster{5}$ indicates this for all subjects, the high importance of $\feature{4}$ indicates that this most often occur for students working with languages, confirming the observations from $\cluster{2}$.
Finally, it is also worth noticing, that there is a strong relation between performance and average session length (clusters $\cluster{1}$, $\cluster{2}$ and $\cluster{3}$), indicating that students, who perform well, also have longer sessions on average.

From the above discussion, it appears that the behavior in clusters $\cluster{4}$ and $\cluster{5}$ are sub-optimal, when considering performance, while students gain more from being in $\cluster{1}$, $\cluster{2}$ or $\cluster{3}$, i.e. by working during school hours, having longer sessions and taking quizzes (in the case of languages) or reading texts (in the case of societal or science subjects).

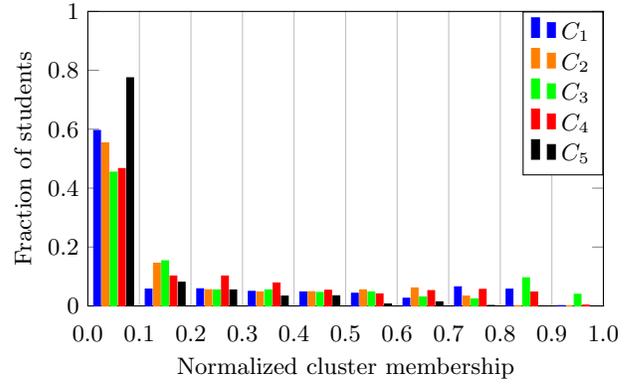
\begin{figure}[t]
    \centering
    \input{figs/e1-dist.tex}
    \caption{The distribution of cluster membership for the first experiment.}
    \label{fig:experiment1-dist}
\end{figure}
\begin{figure}[t]
    \centering
    \input{figs/e1-time.tex}
    \caption{The average cluster membership in each activity period for the first experiment.}
    \label{fig:experiment1-time}
\end{figure}
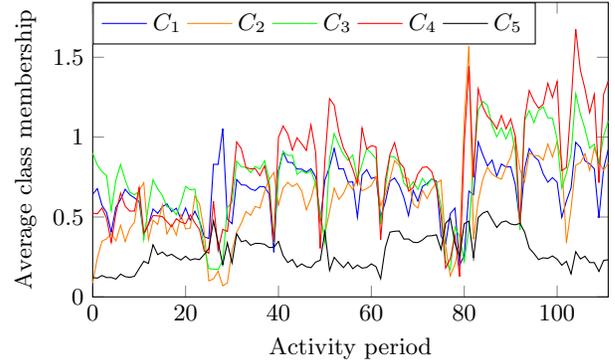

Figure~\ref{fig:experiment1-dist} describes
the distribution of cluster membership across all students and all activity periods
, i.e. the columns of the first interval $[0,0.1)$ gives for each cluster the fraction of students with 0\%-10\% membership.
We see, that we do indeed get a soft clustering, with students often belonging to more than one cluster. Only $\cluster{3}$ seems to be the single dominant cluster of some students. From the figure, we also see that students are typically never exclusively in $\cluster{5}$, which is positive, as the behavior observed in that cluster was not very productive in terms of performance. Other than that, we generally observe that students seem to distribute fairly well between the top four clusters, indicating most time spent during school hours and a varied use of both quizzes and texts across all subjects.

Next, we analyze how the membership of different clusters change over time. Figure~\ref{fig:experiment1-time} plots the average membership for each period, i.e. the average of rows from $\mat{U}$ belonging to the given period.
The first observation we make from Figure~\ref{fig:experiment1-time}, is that clusters $\cluster{1}$, $\cluster{2}$, $\cluster{3}$ and $\cluster{4}$ appear correlated at the system-wide level. This is due to these clusters being dependent on the general activity in the online system; most of the sudden drops
%occurring in Figure~\ref{fig:experiment1-time},
occur at the same time as Danish school vacations, most notably the two larger drops around activity periods 25 and 79 (see Figure~\ref{fig:student-activity}). $\cluster{5}$ seems to be relatively unaffected by the general activity, but this makes sense, as $\cluster{5}$ contains mostly students, who work outside school hours, and thus a lower membership is expected in that cluster in general, which is also the pattern we see in periods with no vacation.

Looking at the general distribution between the different clusters, $\cluster{3}$ and $\cluster{4}$ seem to be the most dominant, indicating that most students are working with language and societal subjects and reading texts. Cluster $\cluster{1}$ (science subjects) is fairly constant in the non-vacation periods, and $\cluster{2}$ seems to increase starting period 80, indicating that more students spend time taking quizzes. Finally, as mentioned, $\cluster{5}$ is the least active cluster across most periods.
One general trend for the top four clusters seem to be an increase in activity during the 112 periods, indicating that students are spending more time in the system on average.

\subsection{Subject and Exercise Complexity}
In the second experiment we look at the relation between subjects and exercises grouped by Bloom's taxonomy level, i.e. we consider features $f_6,f_7,f_8,f_{11},f_{12},f_{13},f_{14}$

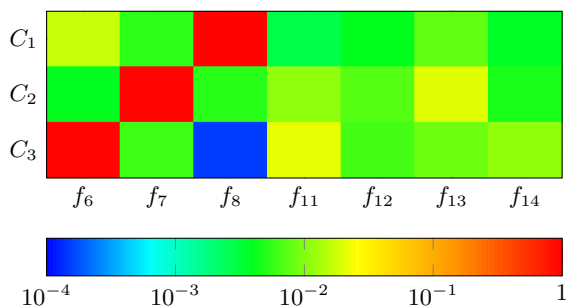
\begin{figure}[t]
    \centering
    \input{figs/e2-V.tex}
    \caption{The cluster matrix for the second experiment. Note, that a logarithmic scale is used for this plot.}
    \label{fig:experiment2-V}
\end{figure}

We expect three clusters, one for each of the subject classes, which will tell us how much each Bloom level is used within each subject class. Figure~\ref{fig:experiment2-V} shows the cluster matrix found.
From Figure~\ref{fig:experiment2-V}, we make the following observations:
\begin{itemize}[noitemsep]
    \item[$\cluster{1}$] In the science subjects, only very little of the 3 higher levels are used, and almost none of reading and understanding.
    \item[$\cluster{2}$] For societal subjects, students have only little activity in the first 2 levels, a lot in analyzing and evaluating, and very little activity in creation.
    \item[$\cluster{3}$] In languages, students have a tendency to read and understand a lot, and then distribute almost evenly on the 3 higher levels.
\end{itemize}
 This implies that if we want to attract students to use an online educational system for languages, focus should be on exercises with Bloom's taxonomy level read and understand. For societal subjects the focus should be on exercises with analyzing and evaluating. For science we see no preference.

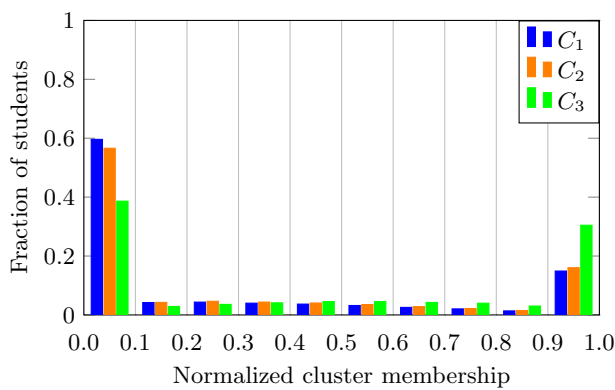
\begin{figure}[t]
    \centering
    \input{figs/e2-dist.tex}
    \caption{The distribution of cluster membership for the second experiment.}
    \label{fig:experiment2-dist}
\end{figure}
From Figure~\ref{fig:experiment2-dist}, we see that the clustering has many high values which is most likely explained by having a teacher who uses the system exclusively in only one of the subjects, which we can see happens most often for languages.

\begin{figure}
    \centering
    \input{figs/e2-time.tex}
    \caption{The average cluster membership in each activity period for the second experiment.}
    \label{fig:experiment2-time}
\end{figure}
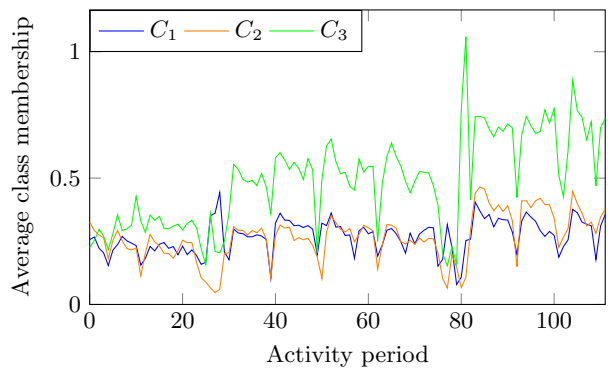

As we can see in Figure~\ref{fig:experiment2-time} all three clusters share similar curvature, which is partly explained by holidays. Especially the science and societal clusters behave seem highly correlated on a general level. We also see that in all three subjects, the average time spent during a week
has gone from 15 minutes, to 45 minutes for languages and 25 minutes for both societal subjects and sciences. A clear indication that teachers and students in Denmark are using online educational systems more, especially for languages.

\section{Conclusions and future work}
Several points can be taken from our analysis.
%% Experiment 1
We have identified three optimal and two sub-optimal behaviors in relation to subject and performance. One notably conclusion is that students using the Clio Online system during non-school hours (at home) do not seem to gain any significant boost to performance. We also saw how taking quizzes seems to increase the performance of students in languages, more so than in other subjects, where reading texts are of more importance. This fits the intuition that skills such as grammar need to be trained, in order to be learned.
%% Experiment 2
We inform how exercises are used depending both on their subject and their level in Bloom's taxonomy.
And lastly we see that the average amount of time spent in the system is increasing both generally and for the individual students in all subjects, but especially for students working with languages.
%% Both
Furthermore, both experiments show how behaviors can have high correlation on a system-wide level, despite being uncorrelated on the individual student level.
While the change of behavior for individual students was not directly analyzed in this paper (due to privacy concerns), our method allows for tracking such individual changes, hopefully helping teachers encourage optimal student behavior, e.g. by recommending training quizzes for students working with languages, or making sure that students are allowed more time to use the system in school.

In our setting, the number of clusters is fixed. It may be interesting to use an adaptive clustering strategy instead, as done in \cite{kli16b}, as one might expect clusters to change over time. In the future, it might also be interesting to include other features, that were not available to us at this time, for instance whether a text (or quiz) have been assigned by a teacher, or whether the student reads it by themselves. For this study, we also only had access to a limited amount of data; better and more reliable results might be obtained by including more data.

\section{Acknowledgments}
The work is supported by the Innovation Fund Denmark through the Danish Center for Big Data Analytics Driven Innovation (DABAI) project. The authors would like to thank Clio Online, and the reviewers for their thorough and insightful feedback.

\bibliographystyle{plain}
\bibliography{bibl}  
\end{document}

%% file: figs/activity.tex
\begin{tikzpicture}

  \begin{axis}[
    ylabel=\#students,
    xlabel=Activity period,
    width=\linewidth,
    height=62mm,
    xmin=0,xmax=111,
    ymin=0,ymax=10000,
    ]
    \addplot[ybar interval,blue,fill=blue] table[x=i,y=students] {figs/data/activity.dat};
  \end{axis}

\end{tikzpicture}

%% file: figs/nmf-cluster.tex
\begin{tikzpicture}
\begin{axis}[
    name=X,
    title={$\mat{X}$},
    height=3.4cm,width=3cm,
    y dir=reverse,
    enlargelimits=false,
    axis on top,
    point meta min=0.0,
    point meta max=0.8,
    xtick=data,
    xticklabels={$\feature{1}$,$\feature{2}$,$\feature{3}$},
    ytick=data,
    yticklabels={$x_1$,$x_2$,$x_3$,$x_4$},
    major tick length=0,
    colorbar,
    colorbar style={
            yticklabel style={
                /pgf/number format/.cd,
                fixed,
                precision=1,
                fixed zerofill,
            },
            ytick={0,0.5,1.0},
            at={(7cm,0cm)},
            anchor=south east
        },
    ]
    \addplot [matrix plot*,point meta=explicit] file [meta=index 2] {figs/data/fig-X.dat};
\end{axis}

\node at (1.95cm,0.9cm) {\LARGE{$\simeq$}};

\begin{axis}[
    name=U,
    title={$\mat{U}$},
    at={($(X.east)+(15mm,0)$)},
    anchor=west,
    height=3.4cm,width=2.6cm,
    y dir=reverse,
    enlargelimits=false,
    axis on top,
    point meta min=0.0,
    point meta max=0.8,
    xtick=data,
    xticklabels={$\cluster{1}$,$\cluster{2}$},
    ytick=data,
    yticklabels={$x_1$,$x_2$,$x_3$,$x_4$},
    major tick length=0,
    ]
    \addplot [matrix plot*,point meta=explicit] file [meta=index 2] {figs/data/fig-U.dat};
\end{axis}

\begin{axis}[
    name=V,
    title={$\mat{V}$},
    at={($(U.east)+(7mm,0)$)},
    anchor=west,
    height=2.6cm,width=3cm,
    y dir=reverse,
    enlargelimits=false,
    axis on top,
    point meta min=0.0,
    point meta max=0.8,
    xtick=data,
    xticklabels={$\feature{1}$,$\feature{2}$,$\feature{3}$},
    ytick=data,
    yticklabels={$\cluster{1}$,$\cluster{2}$},
    major tick length=0,
    ]
    \addplot [matrix plot*,point meta=explicit] file [meta=index 2] {figs/data/fig-V.dat};
\end{axis}
\end{tikzpicture}

%% file: figs/e1-V.tex
\begin{tikzpicture}
    \begin{axis}[
        %view={0}{90},   % not needed for `matrix plot*' variant
        width=\linewidth,
        height=38mm,
        %xlabel=Feature,
        %ylabel=Cluster,
        y dir=reverse,
        colorbar,
        colorbar horizontal,
        %title=Clusters,
        xtick=data,
        xticklabels={$\feature{1}$,$\feature{2}$,$\feature{3}$,$\feature{4}$,$\feature{5}$,$\feature{6}$,$\feature{7}$,$\feature{8}$,$\feature{9}$,$\feature{10}$,$\feature{11}$},
        ytick=data,
        yticklabels={$\cluster{1}$,$\cluster{2}$,$\cluster{3}$,$\cluster{4}$,$\cluster{5}$},
        major tick length=0,
        enlargelimits=false,
        axis on top,
        %point meta min=0.0,
        %point meta max=1.0,
        ]
        \addplot [matrix plot*,point meta=explicit] file [meta=index 2] {figs/data/e1-V.dat};
    \end{axis}
\end{tikzpicture}

%% file: figs/e1-dist.tex
\begin{tikzpicture}

  \begin{axis}[
    legend style={
        at={(1,1)},
        anchor=north east,
        %legend columns=-1
    },
    xlabel={Normalized cluster membership},
    ylabel={Fraction of students},
    width=\linewidth,
    height=55mm,
    xmin=-0.5,xmax=9.5,
    ymin=0,ymax=1,
    ybar=0.2mm,
    bar width=0.9mm,
    xtick={-0.5,0.5,1.5,2.5,3.5,4.5,5.5,6.5,7.5,8.5,9.5,10.5},
    xtick style={
        /pgfplots/major tick length=0pt,
    },
    xticklabels={0.0,0.1,0.2,0.3,0.4,0.5,0.6,0.7,0.8,0.9,1.0},
    xmajorgrids,
    ]
    \addplot[blue,fill=blue] table[x=idx,y=k1] {figs/data/e1-stats.dat};
    \addplot[orange,fill=orange] table[x=idx,y=k2] {figs/data/e1-stats.dat};
    \addplot[green,fill=green] table[x=idx,y=k3] {figs/data/e1-stats.dat};
    \addplot[red,fill=red] table[x=idx,y=k4] {figs/data/e1-stats.dat};
    \addplot[black,fill=black] table[x=idx,y=k5] {figs/data/e1-stats.dat};
    \legend{$\cluster{1}$,$\cluster{2}$,$\cluster{3}$,$\cluster{4}$,$\cluster{5}$}
  \end{axis}

\end{tikzpicture}

%% file: figs/e1-time.tex
\begin{tikzpicture}

  \begin{axis}[
    legend style={
        at={(0,1)},
        anchor=north west,
        legend columns=-1
    },
    xlabel=Activity period,
    ylabel={Average class membership},
    width=\linewidth,
    height=55mm,
    xmin=0,xmax=111,
    ymin=0,%ymax=12000,
    ]
    \addplot[blue] table[x=idx,y=k1] {figs/data/e1-all-students.dat};
    \addplot[orange] table[x=idx,y=k2] {figs/data/e1-all-students.dat};
    \addplot[green] table[x=idx,y=k3] {figs/data/e1-all-students.dat};
    \addplot[red] table[x=idx,y=k4] {figs/data/e1-all-students.dat};
    \addplot[black] table[x=idx,y=k5] {figs/data/e1-all-students.dat};
    \legend{$\cluster{1}$,$\cluster{2}$,$\cluster{3}$,$\cluster{4}$,$\cluster{5}$}
  \end{axis}

\end{tikzpicture}

%% file: figs/e2-V.tex
\begin{tikzpicture}
    \begin{axis}[
        width=\linewidth,
        height=38mm,
        y dir=reverse,
        colorbar,
        colorbar horizontal,
        xtick=data,
        xticklabels={$\feature{6}$,$\feature{7}$,$\feature{8}$,$\feature{11}$,$\feature{12}$,$\feature{13}$,$\feature{14}$},
        ytick=data,
        yticklabels={$\cluster{1}$,$\cluster{2}$,$\cluster{3}$},
        major tick length=0,
        enlargelimits=false,
        axis on top,
        point meta min=-4,
        point meta max=0,
        colorbar style={
            xtick={-4,-3,-2,-1,0},
            xticklabels={$10^{-4}$,$10^{-3}$,$10^{-2}$,$10^{-1}$,$1$}
        },
        ]
        \addplot [matrix plot*,point meta=explicit] file [meta=index 2] {figs/data/e2-V.dat};
    \end{axis}
\end{tikzpicture}

%% file: figs/e2-dist.tex
\begin{tikzpicture}

  \begin{axis}[
    legend style={
        at={(1,1)},
        anchor=north east,
        %legend columns=-1
    },
    xlabel={Normalized cluster membership},
    ylabel={Fraction of students},
    width=\linewidth,
    height=55mm,
    xmin=-0.5,xmax=9.5,
    ymin=0,ymax=1,
    ybar=0.2mm,
    bar width=1.5mm,
    xtick={-0.5,0.5,1.5,2.5,3.5,4.5,5.5,6.5,7.5,8.5,9.5,10.5},
    xtick style={
        /pgfplots/major tick length=0pt,
    },
    xticklabels={0.0,0.1,0.2,0.3,0.4,0.5,0.6,0.7,0.8,0.9,1.0},
    xmajorgrids,
    ]
    \addplot[blue,fill=blue] table[x=idx,y=k1] {figs/data/e2-stats.dat};
    \addplot[orange,fill=orange] table[x=idx,y=k2] {figs/data/e2-stats.dat};
    \addplot[green,fill=green] table[x=idx,y=k3] {figs/data/e2-stats.dat};
    \legend{$\cluster{1}$,$\cluster{2}$,$\cluster{3}$}
  \end{axis}

\end{tikzpicture}

%% file: figs/e2-time.tex
\begin{tikzpicture}

  \begin{axis}[
    legend style={
        at={(0,1)},
        anchor=north west,
        legend columns=-1
    },
    xlabel=Activity period,
    ylabel={Average class membership},
    width=\linewidth,
    height=55mm,
    xmin=0,xmax=111,
    ymin=0,%ymax=12000,
    ]
    \addplot[blue] table[x=idx,y=k1] {figs/data/e2-all-students.dat};
    \addplot[orange] table[x=idx,y=k2] {figs/data/e2-all-students.dat};
    \addplot[green] table[x=idx,y=k3] {figs/data/e2-all-students.dat};
    \legend{$\cluster{1}$,$\cluster{2}$,$\cluster{3}$}
  \end{axis}

\end{tikzpicture}

%% file: main.bbl
\begin{thebibliography}{10}

\bibitem{nmf-overview}
Michael~W. Berry, Murray Browne, Amy~N. Langville, V.~Paul Pauca, and Robert~J.
  Plemmons.
\newblock {Algorithms and Applications for Approximate Nonnegative Matrix
  Factorization}.
\newblock {\em Computational Statistics \& Data Analysis}, 52(1):155 -- 173,
  2007.

\bibitem{conf/edm/FauconKD16}
Louis Faucon, Lukasz Kidzinski, and Pierre Dillenbourg.
\newblock {Semi-Markov model for simulating MOOC students.}
\newblock In {\em Proceedings of the 9th International Conference on
  Educational Data Mining (EDM)}, pages 358--363. International Educational
  Data Mining Society (IEDMS), 2016.

\bibitem{DBLP:conf/edm/GelmanRDVJ16}
Ben~U. Gelman, Matt Revelle, Carlotta Domeniconi, Kalyan Veeramachaneni, and
  Aditya Johri.
\newblock {Acting the Same Differently: A Cross-Course Comparison of User
  Behavior in MOOCs}.
\newblock In {\em Proceedings of the 9th International Conference on
  Educational Data Mining (EDM)}, pages 376--381. International Educational
  Data Mining Society {(IEDMS)}, 2016.

\bibitem{DBLP:journals/corr/abs-1708-04164}
Christian Hansen, Casper Hansen, Niklas Hjuler, Stephen Alstrup, and Christina
  Lioma.
\newblock Sequence modelling for analysing student interaction with educational
  systems.
\newblock In {\em Proceedings of the 10th International Conference on
  Educational Data Mining (EDM)}, pages 232--237. International Educational
  Data Mining Society {(IEDMS)}, 2017.

\bibitem{DBLP:conf/edm/HuttMWDD16}
Stephen Hutt, Caitlin Mills, Shelby White, Patrick~J. Donnelly, and Sidney~K.
  D'Mello.
\newblock {The Eyes Have It: Gaze-based Detection of Mind Wandering during
  Learning with an Intelligent Tutoring System}.
\newblock In {\em Proceedings of the 9th International Conference on
  Educational Data Mining (EDM)}, pages 86--93. International Educational Data
  Mining Society {(IEDMS)}, 2016.

\bibitem{wnmf-overview}
Yong-Deok Kim and Seungjin Choi.
\newblock {Weighted Nonnegative Matrix Factorization}.
\newblock In {\em Proceedings of IEEE International Conference on Acoustics,
  Speech and Signal Processing (ICASSP)}, pages 1541--1544, 2009.

\bibitem{kli16b}
Severin Klingler, Tanja K{\"a}ser, Barbara Solenthaler, and Markus Gross.
\newblock {Temporally Coherent Clustering of Student Data}.
\newblock In {\em Proceedings of the 9th International Conference on
  Educational Data Mining (EDM)}, pages 102--109. International Educational
  Data Mining Society {(IEDMS)}, 2016.

\bibitem{nmf-image-clustering}
Cosmin Lazar and Andrei Doncescu.
\newblock {Non Negative Matrix Factorization Clustering Capabilities;
  Application on Multivariate Image Segmentation}.
\newblock In {\em Proceedings of the 3rd International Conference on Complex,
  Intelligent and Software Intensive Systems (CISIS)}, pages 924--929, 2009.

\bibitem{Lee00}
Daniel~D. Lee and H.~Sebastian Seung.
\newblock {Algorithms for Non-negative Matrix Factorization}.
\newblock In {\em Advances in Neural Information Processing Systems 13, Papers
  from Neural Information Processing Systems {(NIPS)} 2000, Denver, CO, {USA}},
  pages 556--562, 2000.

\bibitem{Li13}
Tao Li and Chris Ding.
\newblock {Non-negative matrix factorization for clustering: A survey}.
\newblock In {\em Data Clustering: Algorithms and Applications}, pages
  149--176. Chapman \& Hall/CRC, January 2013.

\bibitem{Lin07}
Chih-Jen Lin.
\newblock {On the Convergence of Multiplicative Update Algorithms for
  Non-negative Matrix Factorization}.
\newblock {\em Trans. Neur. Netw.}, 18(6):1589--1596, 2007.

\bibitem{ECEL-nmf}
Stephan Lorenzen, Ninh Pham, and Stephen Alstrup.
\newblock {On Predicting Student Performance Using Low-rank Matrix
  Factorization Techniques}.
\newblock In {\em Proceedings of the 16th European Conference on e-Learning
  (ECEL)}, pages 326--334. Academic Conferences and Publishing International,
  2017.

\bibitem{nmf-doc-clustering}
Farial Shahnaz, Michael~W. Berry, Victor~P. Pauca, and Robert~J. Plemmons.
\newblock {Document clustering using nonnegative matrix factorization}.
\newblock {\em Information Processing \& Management}, 42(2):373 -- 386, 2006.

\end{thebibliography}
